\documentclass[aip,jcp,reprint,floatfix]{revtex4-1}
\usepackage[utf8]{inputenc}
\usepackage{amsmath}
\usepackage{amsfonts}
\usepackage{amssymb}

\usepackage{graphicx}
\usepackage{gensymb}
\usepackage{color}
\usepackage{epstopdf}

\begin{document}
\title{Two-structure thermodynamics for the TIP4P/2005 model of water\\ covering supercooled and deeply stretched regions}

\author{John W. Biddle}
\altaffiliation[Now at: ]{Department of Systems Biology, Harvard Medical School, Boston, Massachusetts 02115, USA}
\affiliation{Institute of Physical Science and Technology and Department of Chemical and Biomolecular Engineering, University of Maryland, College Park, Maryland 20742, USA}
\author{Rakesh S. Singh}
\altaffiliation[Now at: ]{Department of Chemistry, Johns Hopkins University, Baltimore, Maryland 21218, USA}
\affiliation{Department of Chemical and Biological Engineering,\\ Princeton University, Princeton, New Jersey 08544, USA}
\author{Evan M. Sparano}
\affiliation{Department of Chemical and Biological Engineering,\\ Princeton University, Princeton, New Jersey 08544, USA}
\author{Francesco Ricci}
\affiliation{Department of Chemical and Biological Engineering,\\ Princeton University, Princeton, New Jersey 08544, USA}
\author{Miguel A. Gonz\'{a}lez}
\altaffiliation[Now at: ]{Department of Chemistry, Imperial College London, London SW7 2AZ, United Kingdom}
\affiliation{Departamento Qu\'{i}mica F\'{i}sica I, Facultad Ciencias Qu\'{i}micas, Universidad Complutense de Madrid,\\
28040 Madrid, Spain}
\author{Chantal Valeriani}
\affiliation{Departamento Qu\'{i}mica F\'{i}sica I, Facultad Ciencias Qu\'{i}micas, Universidad Complutense de Madrid,\\
28040 Madrid, Spain}
\author{Jos\'{e} L. F. Abascal}
\affiliation{Departamento Qu\'{i}mica F\'{i}sica I, Facultad Ciencias Qu\'{i}micas, Universidad Complutense de Madrid,\\
28040 Madrid, Spain}
\author{Pablo G. Debenedetti}
\affiliation{Department of Chemical and Biological Engineering,\\ Princeton University, Princeton, New Jersey 08544, USA}
\author{Mikhail A. Anisimov}
\affiliation{Institute of Physical Science and Technology and Department of Chemical and Biomolecular Engineering, University of Maryland, College Park, Maryland 20742, USA}
\affiliation{Oil and Gas Research Institute of the Russian Academy of Sciences,\\ Moscow 119333, Russia}
\author{Fr\'{e}d\'{e}ric Caupin}\email{frederic.caupin@univ-lyon1.fr}
\affiliation{Univ Lyon, Université Claude Bernard Lyon 1, CNRS, \\ Institut Lumière Matière, F-69622, Villeurbanne, France}

\date{\today}

\begin{abstract}

One of the most promising frameworks for understanding the anomalies of cold and supercooled water postulates the existence of two competing, interconvertible local structures. If the non-ideality in the Gibbs energy of mixing overcomes the ideal entropy of mixing of these two structures, a liquid-liquid phase transition, terminated at a liquid-liquid critical point, is predicted. Various versions of the “two-structure equation of state" (TSEOS) based on this concept have shown remarkable agreement with both experimental data for metastable, deeply supercooled water and simulations of molecular water models. However, existing TSEOSs were not designed to describe the negative pressure region and do not account for the stability limit of the liquid state with respect to the vapor. While experimental data on supercooled water at negative pressures may shed additional light on the source of the anomalies of water, such data are very limited. To fill this gap, we have analyzed simulation results for TIP4P/2005, one of the most accurate classical water models available. We have used recently published simulation data, and performed additional simulations, over a broad range of positive and negative pressures, from ambient temperature to deeply supercooled conditions. We show that, by explicitly incorporating the liquid-vapor spinodal into a TSEOS, we are able to match the simulation data for TIP4P/2005 with remarkable accuracy. In particular, this equation of state quantitatively reproduces the lines of extrema in density, isothermal compressibility, and isobaric heat capacity. Contrary to an explanation of the thermodynamic anomalies of water based on a “retracing spinodal”, the liquid-vapor spinodal in the present TSEOS continues monotonically to lower pressures upon cooling, influencing but not giving rise to density extrema and other thermodynamic anomalies. 
\end{abstract}

\maketitle

\section{Introduction}

The most well-known thermodynamic anomaly of water is the density maximum with respect to temperature, occurring at atmospheric pressure at about 4~\degree C.~\cite{Debenedetti_2003b} Upon supercooling, the behavior of water becomes even more anomalous: density continues to decrease,~\cite{Hare_1987} while the isothermal compressibility and isobaric heat capacity increase sharply.~\cite{Speedy_1976,Angell_1973,Archer_2000} As the pressure is increased, the temperature of maximum density (TMD) along isobars decreases.~\cite{Debenedetti_2003b} One influential hypothesis that explains the anomalous thermodynamics of supercooled water posits the existence of a first-order liquid-liquid phase transition (LLPT) in deeply supercooled water, terminating at a liquid-liquid critical point (LLCP),~\cite{Poole_1992,Mishima_1998b,Stanley_2013,Gallo_2016} in a region where the metastable liquid is difficult to access experimentally due to rapid homogeneous nucleation of ice~.\cite{Debenedetti_2003b}

This hypothesis is consistent with a view that considers water as a “mixture” of two distinct interconvertible local structures: a high-density, high-entropy structure (``structure A") and a low-density, low-entropy structure (``structure B").~\cite{Tanaka_2000b,Bertrand_2011,Holten_2012b,Russo_2014} Structure A is prevalent at high temperatures and pressures, whereas structure B is prevalent at low temperatures and pressures. Based on the two-structure concept, an explicit two-structure equation of state (TSEOS) was developed. Several versions of the TSEOS were successfully used for the description of the thermodynamic anomalies in supercooled water,~\cite{Holten_2012b,Holten_2012a} as well as in different models of water: mW,~\cite{Holten_2013a} ST2,~\cite{Holten_2014JCP} and TIP4P/2005.~\cite{Singh_2016} Sufficient non-ideality in the mixing of these two alternative structures could lead to a liquid-liquid phase transition (as in ST2~\cite{Holten_2014JCP,Sciortino_2011,Palmer_2014,Smallenburg_2015} and, possibly, TIP4P/2005.~\cite{Singh_2016})\nocite{Moore_2009} The existence of a low-density, low entropy structure accounts for the density anomaly upon cooling, as well as for the increase in compressibility and isobaric heat capacity. If there is a liquid-liquid phase transition, then the response functions pass through finite maxima upon isobaric cooling in the one-phase region, with the loci of maxima converging with the critical isochore at the critical point, where the response functions diverge. However,the conjecture of two local structures does not necessarily require that there be a liquid-liquid phase transition, and if such a transition is not present (for example in the mW,~\cite{Holten_2013a} model) then the response functions pass through finite maxima and never diverge.

Experiments~\cite{Taschin_2013,Tokushima_2008,Nilsson_2015} and simulations~\cite{Russo_2014,Singh_2016,Moore_2009} support the existence of two distinct, interconvertible local structures in cold and supercooled water, as well as in water-like models. In particular, the TIP4P/2005,~\cite{Russo_2014,Singh_2016} TIP5P~\cite{Russo_2014} and mW~\cite{Moore_2009} models of water show an increase in the number of molecules with four nearest neighbors and in local tetrahedral arrangements
upon cooling. This is in quantitative agreement with the behavior
of the structure B fraction, the “reaction coordinate” in
two-structure thermodynamics.

Recent experimental progress has revived interest in the doubly metastable region, where liquid water is both supercooled and under tension.~\cite{Caupin_2015} The doubly metastable region was explored as a novelty by Hayward in 1971.~\cite{Hayward_1971} Subsequent experiments~\cite{Henderson_1980,Zheng_1991,Herbert_2006} further explored this region. The recent experiments of Pallares et al~\cite{Pallares_2014,Pallares_2016}. accomplished a significant penetration into the region where water is metastable with respect to both the crystal and vapor phases. In particular, the line of density maxima was investigated down to -120 MPa.~\cite{Pallares_2016} Although the available experimental data are still sparse, they reveal the inadequacy of extrapolations of positive-pressure behavior into the negative-pressure region. For example, the speed of sound can reach nearly twice the value predicted by extrapolation.~\cite{Pallares_2014} 

Liquid water cannot be stretched indefinitely. Eventually, even the metastability of the liquid state must end at the liquid-vapor spinodal (LVS), the absolute stability limit of liquid with respect to the vapor. At the LVS, the isothermal compressibility diverges, as does the isobaric heat capacity. None of the previous versions of the TSEOS account for the existence of the liquid-vapor spinodal and, accordingly, none has been used for the study of negative pressures. This was a serious limitation in the applications of two-structure thermodynamics, especially in view of the fact that the shape of the spinodal and its possible connection to supercooled water’s anomalies have been debated since the 1980s.~\cite{Debenedetti_2003b,Speedy_1982,Debenedetti_1986,Poole_1993,Poole_1994,Sastry_1996,Speedy_2004,Debenedetti_2004,Stokely_2010,Lu_2016,Angell_2016} In 1982 Speedy~\cite{Speedy_1982} proposed an interpretation of the thermodynamic anomalies of metastable water. He conjectured “that a continuous line of stability limits bounds the superheated, stretched, and supercooled states”, which would cause the increase in response functions upon supercooling. This line of instability is unlikely to be a “retracing liquid-vapor spinodal”, as has been argued on thermodynamic grounds.~\cite{Debenedetti_2003b,Speedy_2004,Debenedetti_2004} However, the debates on the behavior of the stability limits in doubly metastable water are far from over, especially in view of a “critical-point-free” scenario, the possibility of continuation of the first-order LLPT down to the absolute stability of the liquid state.~\cite{Gallo_2016,Poole_1994,Stokely_2010,Angell_2016}

In the present work, we have applied two-structure thermodynamics to the description of recently published~\cite{Singh_2016,Gonzalez_2016,Sumi_2013} and new, previously unpublished extensive simulation data from the Princeton group on the thermodynamic properties of the TIP4P/2005 classical water model over a wide range of temperatures and pressures. TIP4P/2005 is one of the best available models of water, and, in particular, it reproduces well the thermodynamic anomalies of real water at low temperatures,~\cite{Abascal_2005} and the sound velocity in stretched water.~\cite{Pallares_2014} The question of the existence of a LLPT in TIP4P/2005 continues to be debated, with several studies having argued in favor of liquid-liquid separation,~\cite{Sumi_2013,Abascal_2010,Yagasaki_2014} while a recent study reported the disappearance of the transition upon increasing the size of the simulated system.\cite{Overduin_2015} One of the difficulties with low-temperature simulations of this model is the rapid increase of the structural relaxation time in the deeply supercooled region. In any case, the thermodynamic surface of the model evaluated in the one-phase region above 180 K clearly shows the hallmarks of criticality (at about 182 K, 170 MPa, and 1017 $\mathrm{kg\,m^{-3}}$), while an equation of state based on two-structure thermodynamics shows excellent agreement with the simulation data at positive pressures.~\cite{Singh_2016} In this work, we extend two-structure thermodynamics to negative pressures down to the liquid-vapor spinodal and explicitly include the spinodal into the analysis. Agreement between the simulation data and the TSEOS is remarkable. The liquid-vapor spinodal significantly affects the thermodynamic behavior of the model in the doubly metastable region. However, contrary to an explanation of the thermodynamic anomalies of water based on a “retracing spinodal”, the liquid-vapor spinodal in the TSEOS continues monotonically to lower pressures upon cooling, influencing but not giving rise to density extrema and other thermodynamic anomalies.

\section{Equation of state}

We model the thermodynamic behavior of supercooled water with a TSEOS similar to that used in Refs.~\onlinecite{Holten_2012b,Holten_2013a,Holten_2014JCP,Singh_2016}, but with a significant addition: a liquid-vapor spinodal (LVS) at negative pressures. In keeping with two-structure thermodynamics, water is viewed as a ``mixture'' of two distinct local structures: a high-density, high-entropy structure (``structure A") prevalent at high temperatures, and a low-density, low-entropy structure (``structure B") prevalent at low temperatures. The molar Gibbs energy of this ``mixture" takes the form
\begin{eqnarray}
G & = & G^\mathrm{A} + xG^\mathrm{BA}\notag \\
& + & RT\left[x\mathrm{ln}x + (1 - x)\mathrm{ln}(1 - x) + \omega x (1 - x)\right],
\end{eqnarray}
where $x$ is the fraction of molecules in structure B. $G^\mathrm{A}$ is the Gibbs energy of pure structure A, while $G^\mathrm{BA}$ is the difference in Gibbs energy between structure B and structure A. $\omega$ is a parameter describing the non-ideality of the mixture. $G^\mathrm{A}$, $G^\textrm{BA}$, and $\omega$ are each functions of $T$ and $P$. 

\begin{figure}[tttt]
\includegraphics[width = 0.99\columnwidth]{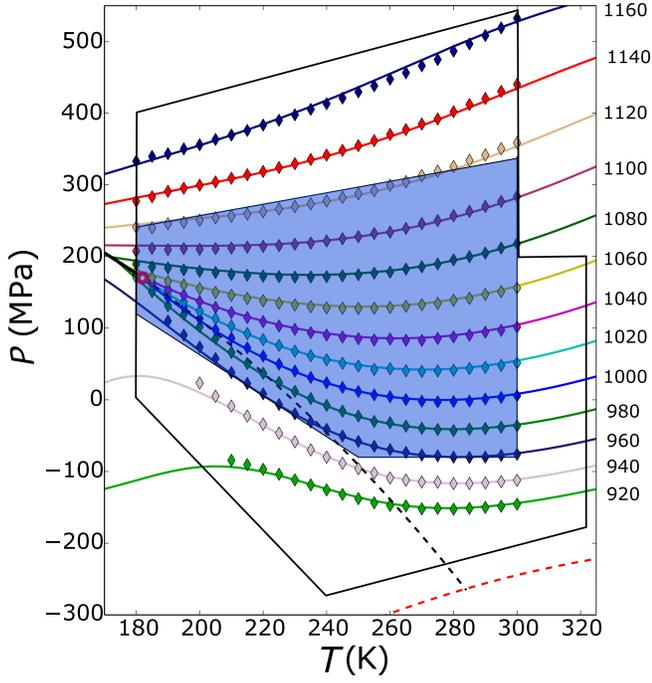}
\caption[Isochores in TIP4P/2005 over broad range]{Pressure along isochores. Symbols show simulations from the Princeton group of various isochores at the densities given on the right-hand side of the graph (in $\mathrm{kg\,m^{-3}}$). Solid lines of the same color represent the predictions of our TSEOS. The solid black line, open red circle, and dashed black line are the LLPT, LLCP, and Widom line, respectively. The dashed red line is the liquid-vapor spinodal. Isochores 960-1120~$\mathrm{kg\,m^{-3}}$ were first published in Ref.~\onlinecite{Singh_2016}, while others are published here for the first time. All the data shown in this figure were included in the fit. The domains of validity of the previous~\cite{Singh_2016} and present (extended) versions of TSEOS are shown with a shaded area and solid-line box, respectively. \label{RhoIsochores}}
\end{figure}

The two structures A and B are interconvertible, so unlike in a binary mixture, the fraction $x$ of molecules participating in structure B is not an independent thermodynamic variable. Rather, it is controlled by the condition of thermodynamic equilibrium
\begin{equation}
\left(\frac{\partial G}{\partial x}\right)_{T,P}= 0.\label{eq:dgdx}
\end{equation}
To find the value of any thermodynamic property with the TSEOS for a given $T$ and $P$, one must first compute the equilibrium fraction $x_\mathrm{e}(T,P)$ from Eq. \eqref{eq:dgdx}, and then evaluate the appropriate derivative of the Gibbs energy at the given conditions $(T,P;x_\mathrm{e})$.

The TSEOS includes a critical point, so it is convenient to work in terms of the reduced variables $\Delta \hat{T} = (T - T_\mathrm{c})/T_\mathrm{c}$ and $\Delta \hat{P} = (P - P_\mathrm{c})/(\rho_\mathrm{c} R T_\mathrm{c})$, where $T_\mathrm{c}$, $\rho_\mathrm{c}$, and $P_\mathrm{c}$ are the critical temperature, density and pressure, respectively. $R$ is the universal gas constant. In general, we work with dimensionless variables, which are reduced by the appropriate combination of $T_\mathrm{c}$, $\rho_\mathrm{c}$, and $R$, \emph{e.\hspace{1mm}g.} $\hat{G} = G/(RT_\mathrm{c})$. 

\begin{figure}[tttt]
\includegraphics[width = 0.99\columnwidth]{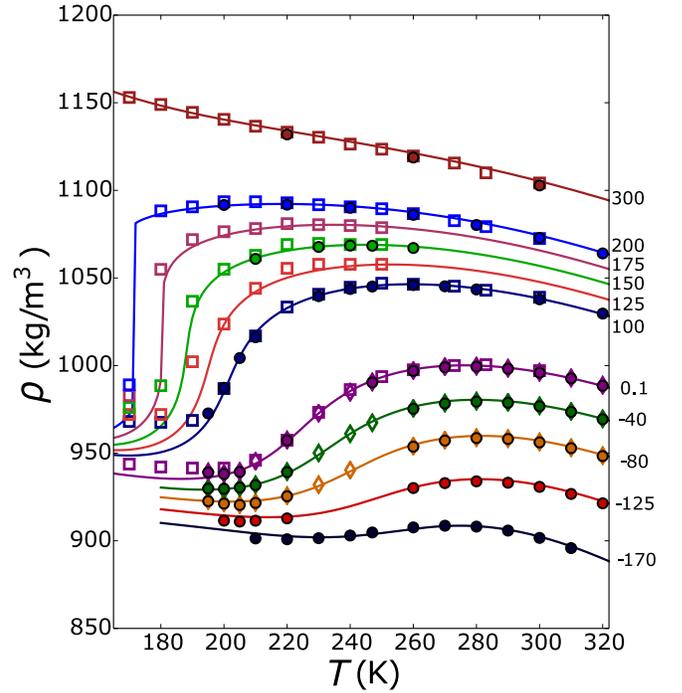}
\caption[Density isobars in TIP4P/2005 from negative to positive pressure]{Density along isobars. Symbols show simulations of various isobars as indicated on the right-hand side of the graph (in $\mathrm{MPa}$). Simulations from the Madrid group~\cite{Gonzalez_2016} included in the TSEOS fit are shown by filled circles. Other data from the Madrid group~\cite{Gonzalez_2016} (open diamonds) and from Ref.~\onlinecite{Sumi_2013} (open squares) were not included in the fit, but are well described by the TSEOS, whose predictions are shown by solid lines of the same color as the symbols.\label{RhoIsobars}}
\end{figure}

\begin{figure*}[tttt]
\includegraphics[width = 0.99\textwidth]{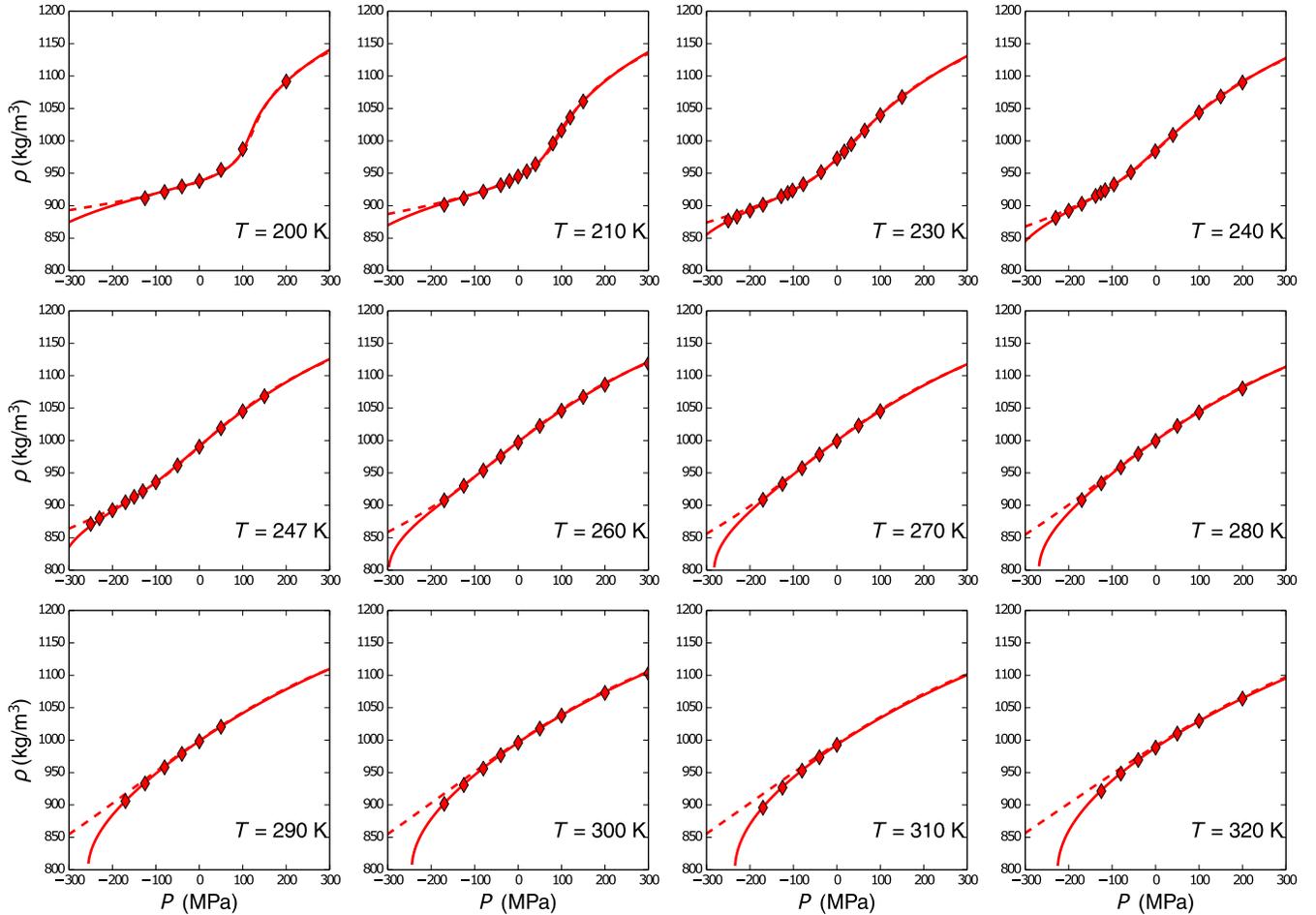}
\caption[Density along isotherms in TIP4P/2005 including negative pressures]{Density along isotherms. Symbols show simulation data from the Madrid group~\cite{Gonzalez_2016} (included in the fit). Solid lines show the TSEOS fitted for this work; dotted lines show the extrapolation of the equation of state presented in Ref.~\onlinecite{Singh_2016}. \label{RhoIsotherms}}
\end{figure*}
%
%
The behavior of pure structure A is represented by
\begin{equation}
\hat{G}^\textrm{A} = \hat{G}^\sigma + \sum_{m,n} c_{mn}\Delta \hat{T}^m \Delta \hat{P}^n \, .
\label{eq:GA}
\end{equation}
The term $\hat{G}^\sigma$ accounts for the effects of the spinodal and will be discussed later. The coefficients $\{c_{mn}\}$ are adjustable parameters to be fitted to the data. In fitting the equation of state, we found it necessary and sufficient to include terms up to fourth order in $\Delta \hat{T}$ and $\Delta \hat{P}$.  We also noticed that the $\Delta \hat{P}^4$ term did not improve the fit and it was therefore discarded. Due to the overall symmetry of the equation of state, the condition $G^\mathrm{BA}=0$ locates the LLPT, the LLCP, and the Widom line, that is, the line of maximum fluctuations of the order parameter that continues the LLPT into the one-phase region. $\hat{G}^\mathrm{BA}$ is therefore expressed as
\begin{equation}
\hat{G}^\textrm{BA} = \lambda (\Delta \hat{T} + a \Delta \hat{P} + b \Delta \hat{T} \Delta \hat{P} + d \Delta \hat{P}^2 + f \Delta \hat{T}^2).
\end{equation}
In this formulation, $\lambda$ and $\lambda a$ give the difference in entropy and volume between structure A and structure B, respectively, at the critical point. $\lambda b$, $\lambda d$, and $\lambda f$ give the corresponding differences in the isobaric expansion coefficient $\alpha_P$, the isothermal compressibility $\kappa_T$, and the isobaric heat capacity $C_P$, respectively. We can find the slope of the LLPT at the critical point as $(d\hat{P}/d\hat{T}) = -1/a$, and the other parameters contribute to the curvature of the LLPT and Widom line.

\begin{figure}[tttt]
\includegraphics[width = 0.99\columnwidth]{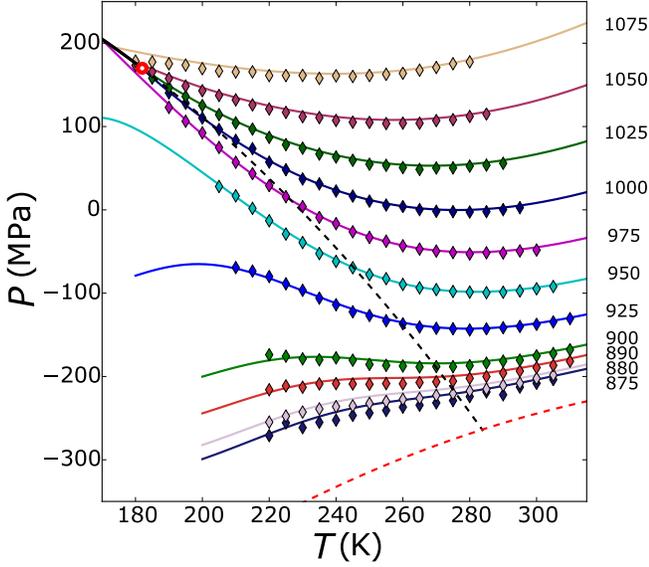}
\caption[Additional isochores including very low densities]{Pressure along isochores. Symbols show new simulations from the Princeton group of various isochores at the densities given on the right-hand side of the graph (in $\mathrm{kg\,m^{-3}}$). Solid lines of the same color represent the predictions of the TSEOS. Although the simulation data shown here were not used in the fitting of the TSEOS, it can be seen that the TSEOS extrapolates well to very low densities and pressures. The solid black line, open red circle, dashed black line, and dashed red line are the LLPT, LLCP, Widom line, and LVS, respectively. \label{SparanoRhoIsochores}}
\end{figure}
The Gibbs energy of mixing is expressed as $RT [x \mathrm{ln} x + (1 - x)\mathrm{ln}(1 - x) $ $ + \omega x (1 - x)]$. The term $RT\left[ x \mathrm{ln} x + (1 - x)\mathrm{ln}(1 - x)\right] $ is the contribution to the Gibbs energy arising from the ideal or Lewis-Randall entropy of mixing associated with a binary solution.~\cite{Rowlinson_1982} We model the non-ideal contribution to the Gibbs energy of mixing in a simple, symmetric form: $RT \omega x (1- x)$, with the parameter $\omega(T,P)$ controlling the magnitude of the non-ideality. Criticality requires $\omega = 2$; for larger values there will be an LLPT, while smaller values indicate the one-phase region where non-ideality is too weak to generate phase separation. We use the form
\begin{equation}
\omega = \frac{2 + \omega_0 \Delta \hat{P}}{\hat{T}}
\end{equation}
for the non-ideality of mixing. With $\omega$ in this form, the TSEOS has no non-ideal entropy of mixing. Consequently, the resulting phase transition has been referred to as an energy-driven LLPT, although it should be noted that non-ideality in both volume and energy of mixing contribute to the phase transition.

We include the liquid-vapor spinodal through the term $G^\sigma$ in Eq.~\ref{eq:GA} with a construction similar to that introduced in Ref.~\onlinecite{Speedy_1982}: because $\left(\partial P/\partial V \right)_{T}$ must vanish at the LVS, $P$ in the vicinity of the LVS can be expanded as a function of $V$ in a Taylor series whose first non-vanishing, non-constant term will be of second order in $V$. To second order, then,
\begin{equation}
P = P_s + \frac{1}{2}\left(\frac{\partial^2 P}{\partial V^2}\right)_{T,P=P_s}(V - V_s)^2,\label{eq:Spinodal1}
\end{equation} 
where $P_s$ and $V_s$ are the pressure and volume of the liquid at the LVS. In this case, the asymptotic behaviors for the volume $V$ and the isothermal compressibility $\kappa_T$ upon approaching the spinodal are
\begin{align}
V_s - V &\sim (P - P_s)^{1/2} \\
\kappa_T &\sim (P - P_s)^{-1/2},
\end{align}
as is predicted, for example, in the classical Van der Waals treatment of the liquid-vapor transition.

The relationship in Eq. \ref{eq:Spinodal1} and the resulting asymptotic behavior can be introduced into the equation of state if the Gibbs energy $G^\textrm{A}$ in Eq.~\ref{eq:GA} includes a term of the form
\begin{equation}
G^\sigma(T,P) = A(T)(P - P_s(T))^{3/2}.
\end{equation}

From this expression we find the contributions (indicated by $\Delta$) of the term $G^\sigma$ to the thermodynamic properties as follows: 
\begin{align}
\Delta V &= \frac{3}{2} A (P - P_\mathrm{s})^{1/2} \label{eq:SpinodalV} \\
\Delta \left(V\kappa_T \right) &= -\frac{3}{4} A (P - P_\mathrm{s})^{-1/2}, \\
\frac{\Delta C_P}{T} &= - \frac{3}{4} A \left(\frac{d P_\mathrm{s}}{d T}\right)^2 (P - P_\mathrm{s})^{-1/2}\notag \\ &+ 3 \left(\frac{d A}{d T}\right) \left(\frac{d P_\mathrm{s}}{d T}\right)(P - P_\mathrm{s})^{1/2}\notag \\
&- \left(\frac{d^2 A}{d T^2}\right)(P - P_\mathrm{s})^{3/2}\notag \\ &+ \frac{3}{2}A\left(\frac{d^2P_\mathrm{s}}{dT^2}\right)(P-P_\mathrm{s})^{1/2}, \\
\Delta \left(V\alpha_P \right) &= -\frac{3}{4}A\left(\frac{dP_\mathrm{s}}{dT}\right)(P - P_\mathrm{s})^{-1/2}\notag \\
&+ \frac{3}{2}\left(\frac{d A}{d T}\right)(P - P_\mathrm{s})^{1/2}.
\end{align}


\begin{table}[tttt]
\centering
\caption{Parameters for the two structure equation of state}
\begin{tabular}{cccc}
\hline
\hline
Parameter	& Value	& Parameter	& Value	\\
$T_\mathrm{c}$	& 182$\,\mathrm{K}$	& $c_{02}$ & -0.00261876	\\
$P_\mathrm{c}$	& 170$\,\mathrm{MPa}$	& $c_{11}$ & 0.257249	\\
$\rho_\mathrm{c}$	& 1017$\,\mathrm{kg\,m^{-3}}$	& $c_{20}$ & -6.30589	\\
$\lambda$ & 1.55607 & $c_{03}$ & 0.000605678	\\
$a$ & 0.154014 & $c_{12}$ & 0.0248091	\\
$b$ & 0.125093 & $c_{21}$ & -0.0400033 \\
$d$ & 0.00854418 & $c_{30}$ & 2.18819 \\
$f$ & 1.14576 &	$c_{13}$ & -0.000994166 \\
$\omega_0$ & 0.03 & $c_{22}$ & -0.00840543 \\
$A_0$ & -0.0547873	& $c_{31}$ & 0.0719058 \\
$A_1$ & -0.0822462	& $c_{40}$ & -0.256674 \\
$S_0$ & -5.40845 &	&	\\
$S_1$ & 5.56087 &	&	\\
$S_2$ & -2.5205 &	&	\\
\hline
\end{tabular}\label{params}
\end{table}

Thus $\kappa_T$ will diverge as $(P - P_\mathrm{s})^{-1/2}$, and, using Eqs. \ref{eq:Spinodal1} and \ref{eq:SpinodalV} we can identify
\begin{equation}
A = -\frac{2\sqrt{2}}{3}\left(\frac{\partial^2P}{\partial V^2}\right)_{T;P=P_\mathrm{s}}^{-1/2}.
\end{equation}
For this work, $\hat{A}(T)$ takes the form
\begin{equation}
\hat{A}(T) = A_0 + A_1\Delta \hat{T},
\end{equation}
where $A_0$ and $A_1$ are optimized to fit the data. 

\begin{figure*}[tttt]
\includegraphics[width = 0.75\textwidth]{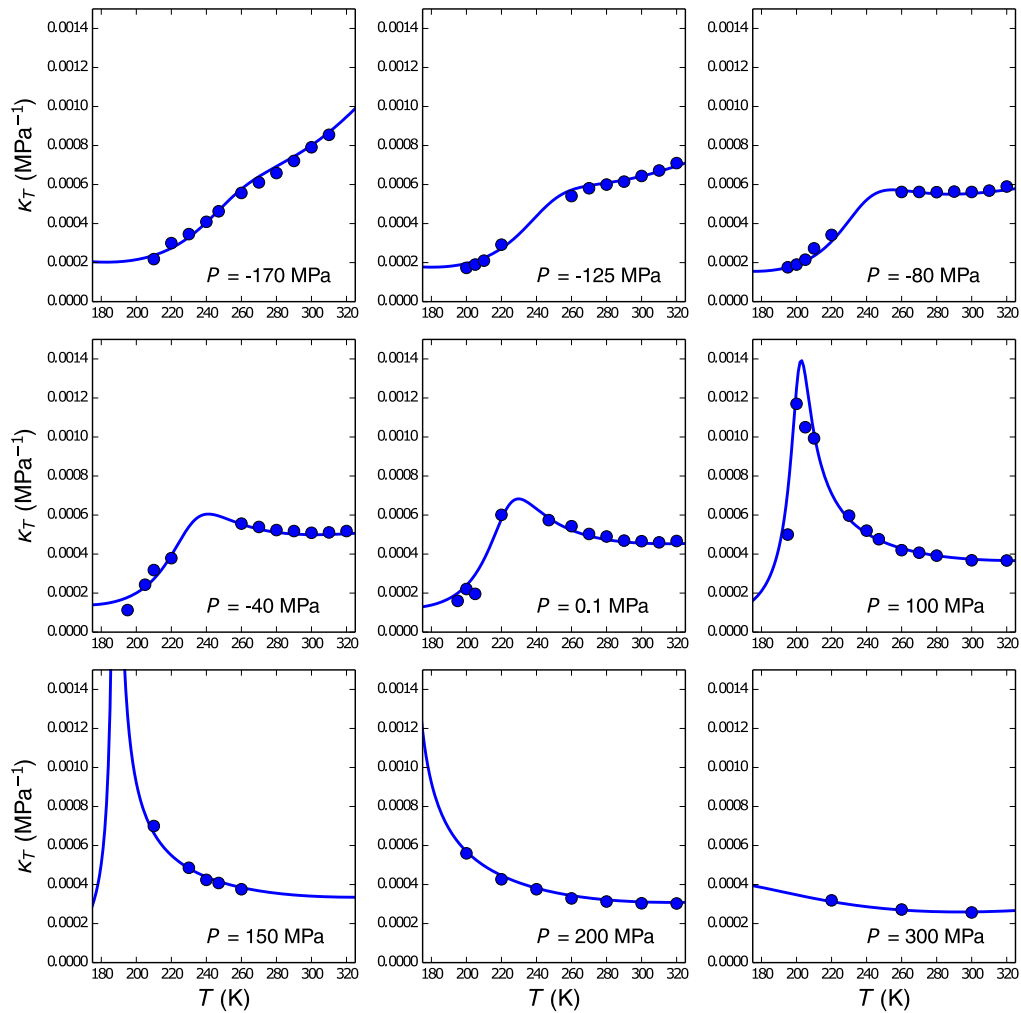}
\caption[Isothermal compressibility along isobars in TIP4P/2005 including negative pressures]{Isothermal compressibility along isobars. Symbols show simulation data from the Madrid group~\cite{Gonzalez_2016} (included in the fit). Solid lines are fits by the TSEOS. \label{KtIsobars}}
\end{figure*}

Where the slope $(dP_\mathrm{s}/dT)$ of the LVS is finite, $C_P$ and $\alpha_P$ diverge with the same exponent as $\kappa_T$, and in the neighborhood of the LVS, $\alpha_P$ must have the same sign as the slope of the LVS. At an extremum in the spinodal pressure as a function of temperature, however, where $(dP_\mathrm{s}/dT) = 0$, $C_P$ and $\alpha_P$ do not diverge. In fact, as Ref.~\onlinecite{Speedy_1982} has shown, $\alpha_P = 0$ at such a point. From these considerations, Ref.~\onlinecite{Speedy_1982} demonstrates that if the the TMD intersects the LVS it must do so at a minimum in the LVS, and that, conversely, if the LVS goes through a minimum, a TMD line must become tangent to the LVS at that minimum. For TIP4P/2005 water, whose TMD locus reaches a maximum temperature and then retraces towards low temperatures upon decreasing the pressure, this means that the points of minimum density that we have observed along isobars are inconsistent with a retracing spinodal. Consequently, we use a monotonic LVS to model TIP4P/2005.




The shape of the LVS, $P_\mathrm{s}(T)$, is represented by a quadratic dependence on temperature:
\begin{equation}
\hat{P}_\mathrm{s}(T) = S_0 + S_1\Delta\hat{T} + S_2 \Delta \hat{T}^2.
\end{equation}

In order to assign values to the parameters $\{S_n\}$, we used the cavitation pressure of TIP4P/2005 along several isotherms, reported in Ref.~\onlinecite{Gonzalez_2016}. We carried out a least-squares fit to these data, and assigned to $S_1$ and $S_2$ exactly the values derived from this least-squares fit. Because one always observes cavitation in simulations before the LVS can be reached,~\cite{Binder_2014} $S_0$ is arbitrarily adjusted
down by a constant downward shift relative to the observed cavitation pressure. Thus the spinodal has the same shape in the $(T,P)$ plane as the simulated cavitation line, but lies at lower pressures. A $-25\,\mathrm{MPa}$ shift was chosen as it gave the best fit results.

\begin{figure*}[tttt]
\includegraphics[width = 0.99\textwidth]{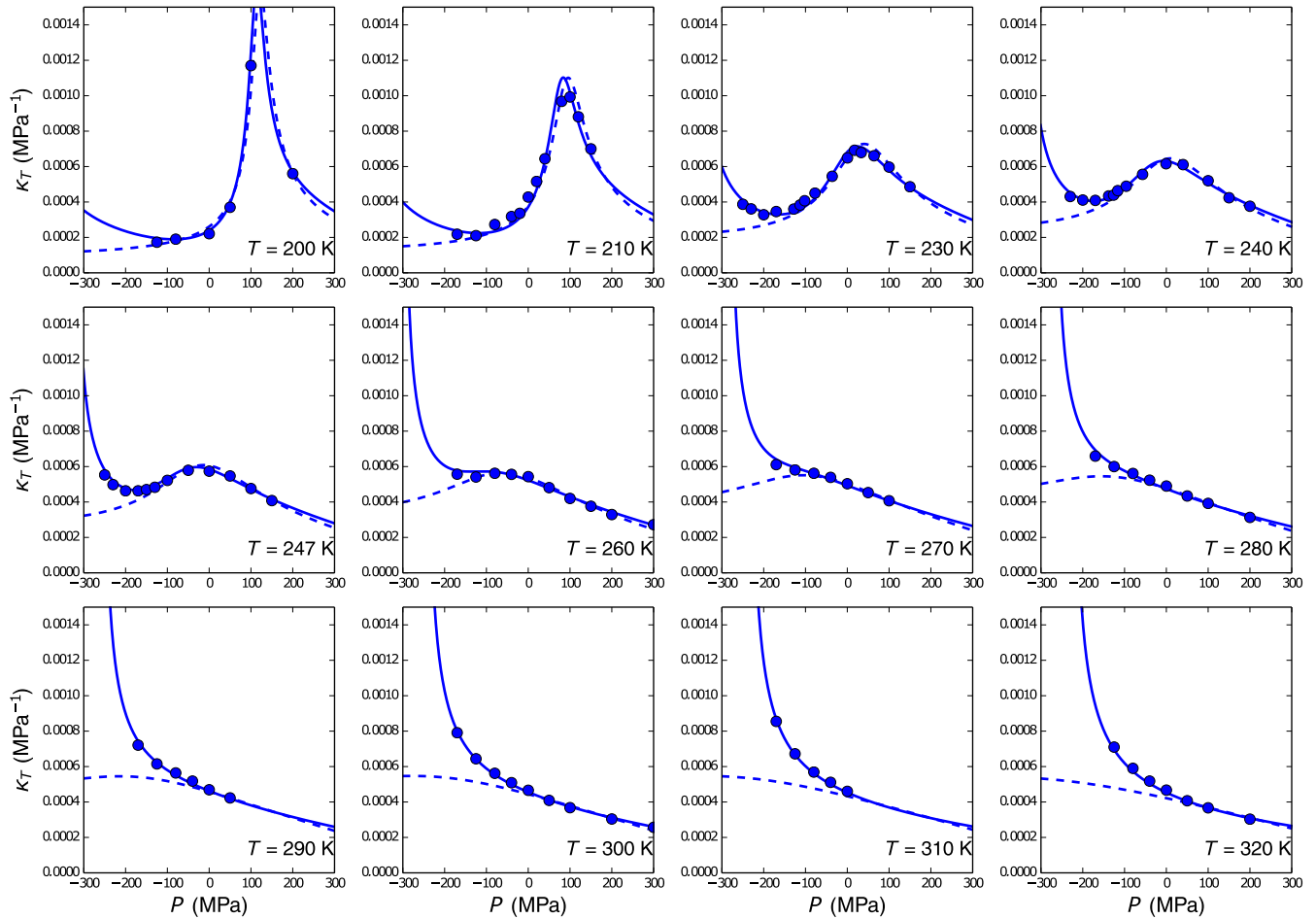}
\caption[Isothermal compressibility along isotherms in TIP4P/2005 including negative pressures]{Isothermal compressibility along isotherms. Symbols show simulation data from the Madrid group~\cite{Gonzalez_2016} (included in the fit). Solid lines show the TSEOS fitted for this work; dotted lines show the extrapolation of the equation of state presented in Ref.~\onlinecite{Singh_2016}. \label{KtIsotherms}}
\end{figure*}

\section{Simulation data\label{sec:simul}}

The TSEOS including a LVS described in the preceding section was fit on selected simulation data of the TIP4P/2005 model. The first set of fitted data is shown in Fig.~\ref{RhoIsochores}. These simulations from the Princeton group were performed along isochores, at densities ranging from 920 to 1160 $\mathrm{kg\,m^{-3}}$ and temperatures ranging from 180 to 300 K, as shown in Fig.~\ref{RhoIsochores}. Note that isochores ranging from 960 to 1120 $\mathrm{kg\,m^{-3}}$ (indicated by the shaded region in Fig.~\ref{RhoIsochores}) were first published in Ref.~\onlinecite{Singh_2016}. 
The second set of fitted data corresponds to Ref.~\onlinecite{Gonzalez_2016}. These simulations from the Madrid group were performed along isobars and isotherms, covering pressures from -170 to 300 MPa and temperatures from 195 to 320 K. Data along isobars and isotherms are displayed in Fig.~\ref{RhoIsobars} (using filled circles), and Fig.~\ref{RhoIsotherms} (using filled diamonds), respectively. In addition to temperature, pressure, and density, the isochoric heat capacity available from the Princeton~\cite{Singh_2016} simulations, and the isothermal compressibility, the isobaric heat capacity, and the isochoric heat capacity available from the Madrid simulations~\cite{Gonzalez_2016} were included in the fit. Simulation details for these two sets of fitted data can be found in Refs.~\onlinecite{Singh_2016} and~\onlinecite{Gonzalez_2016}. While there are slight differences in the technical specifications used for the two sets of simulations, a comparison of the two data sets along the 0.1~MPa isobar shows that they are compatible with each other. For isothermal compressibility any discrepancy between the two data sets is much smaller than the uncertainty associated with the simulations. The discrepancies in the density are on the order of 0.1\%.

\begin{figure*}[tttt]
\includegraphics[width = 0.85\textwidth]{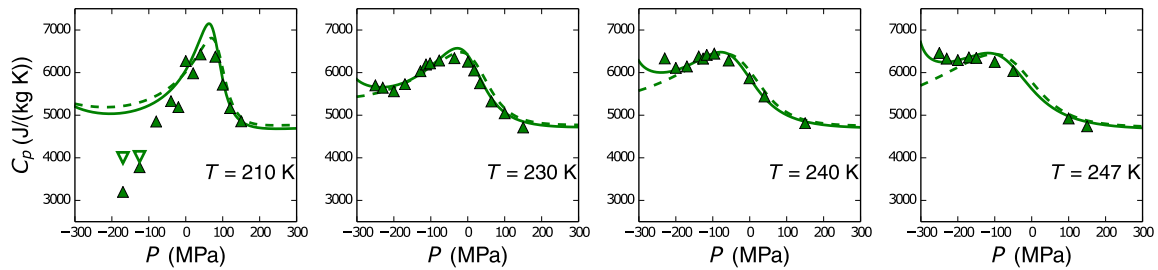}
\caption[Isobaric heat capacity along isotherms in TIP4P/2005 including negative pressures]{Isobaric heat capacity along isotherms. Filled symbols show simulation data from the Madrid group~\cite{Gonzalez_2016} (included in the fit). Solid lines show the TSEOS fitted for this work; dotted lines show the extrapolation of the equation of state presented in Ref.~\onlinecite{Singh_2016}. Open symbols show longer simulation runs ($500\,\mathrm{ns}$) performed separately from the simulations used in the fit.\label{CpIsotherms}}
\end{figure*}

\begin{figure*}[tttt]
\includegraphics[width = 0.75\textwidth]{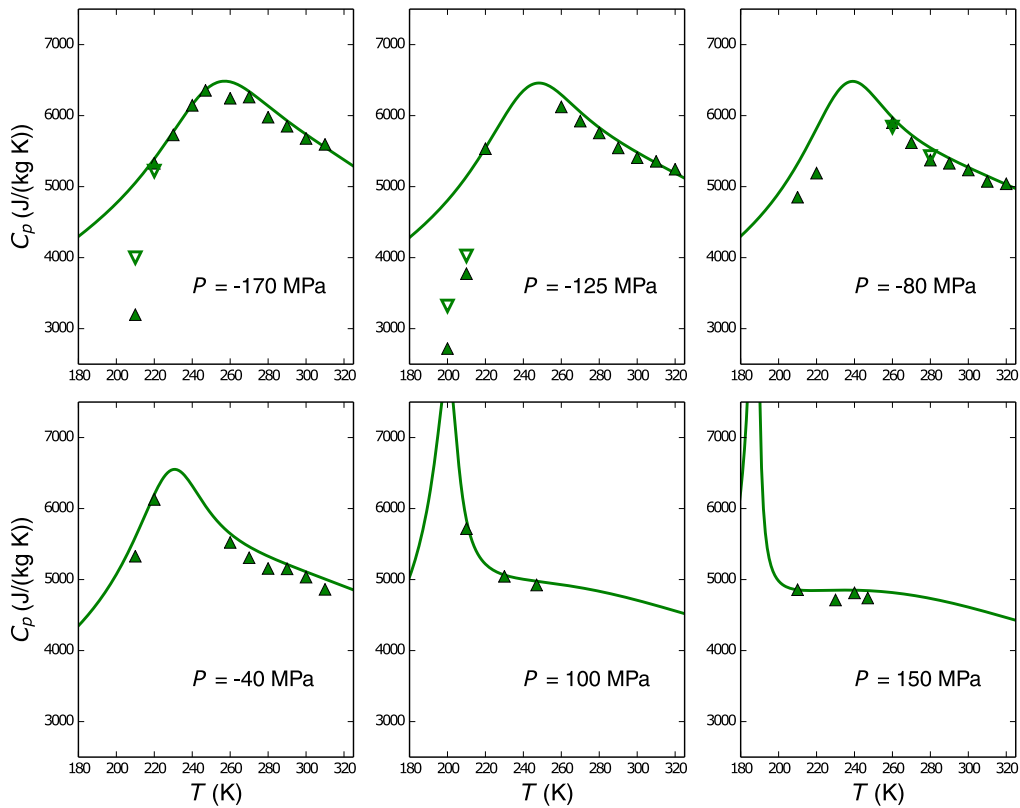}
\caption[Isobaric heat capacity along isobars in TIP4P/2005 including negative pressures]{Isobaric heat capacity along isobars. Filled symbols show simulation data from the Madrid group~\cite{Gonzalez_2016} (included in the fit). Solid lines are fits by the TSEOS. Open symbols show longer simulation runs ($500\,\mathrm{ns}$) performed separately from the simulations used in the fit.\label{CpIsobars}}
\end{figure*}
Other simulation data were not included in the fit by the TSEOS, but are used to test the predictions of the fitted TSEOS. These ``test'' data are made of three sets. The first set corresponds to the remaining simulation data from the Madrid group~\cite{Gonzalez_2016} which were not used in the fit (open diamonds in Fig.~\ref{RhoIsobars}). The second set corresponds to simulations along isobars by Sumi and Sekino~\cite{Sumi_2013} (open squares in Fig.~\ref{RhoIsobars}) which extend to lower temperatures than the fitted data. The third set corresponds to additional simulations along isochores especially performed for the present work by the Princeton group. They are shown in Fig.~\ref{SparanoRhoIsochores}. The new data overlap with the previous data from Princeton,~\cite{Singh_2016} but also include lower density isochores. Simulations were performed on 216 particles systems using the GROMACS 4.5.6 molecular dynamics simulation package. Periodic boundary conditions were applied, and a time step of 1 fs was used. The short-range interactions were truncated at 8.5 Å. Long range electrostatic terms were computed by particle mesh Ewald with a grid spacing 1.2 Å. Long range corrections were applied to the short range Lennard-Jones interaction for both energy and pressure. Bond constraints were maintained using the LINCS (Linear Constraint Solver) algorithm. A Nose-Hoover thermostat with a 1 ps relaxation time was used to maintain constant temperature.

\section{Results and discussion}

The best fit parameters for the TSEOS are given in Table~\ref{params}. The equilibrium fraction $x_\mathrm{e}$ of molecules participating in structure B is shown in Supplementary Figure S1 along selected isobars and isotherms. $x_\mathrm{e}$ increases upon lowering temperature and pressure. Figure~\ref{RhoIsochores} compares $PVT$ data along isochores from simulations by the Princeton group~\cite{Singh_2016} with the corresponding isochores plotted from the fitted TSEOS. The present equation of state matches density data over a broader range than any previous fit, without any sacrifice of quality in the critical region. The TSEOS also matches the density data from the Madrid group~\cite{Gonzalez_2016} which were included in the fit, as can be seen along isobars in Fig.~\ref{RhoIsobars} and along isotherms in Fig.~\ref{RhoIsotherms}. The quality of the TSEOS is further illustrated by the fact that it reproduces well other ``test'' data (see Section~\ref{sec:simul}), which were not included in the fit and extend in a region not covered by the fitted data set. This can be seen in Fig.~\ref{RhoIsobars} for the data from the Madrid group~\cite{Gonzalez_2016} not included in the fit (open diamonds) and for the data from Sumi and Sekino~\cite{Sumi_2013} (open squares). This is also demonstrated by Fig.~\ref{SparanoRhoIsochores} showing the agreement between the TSEOS and the new simulation data from the Princeton group, which extend close to the LVS. The comparison with fitted data and ``test'' data allows us to delimit the region of validity of the present TSEOS which is displayed in Fig.~\ref{RhoIsochores} and extends significantly the region of validity of the previous TSEOS~\cite{Singh_2016} which did not account for the LVS. The pre-spinodal effects are most clearly visible in the behavior of the higher-temperature isotherms, shown in Fig.~\ref{RhoIsotherms} and the TSEOS accounts well for these isotherms. However, the improvement of this version of the TSEOS over previous versions is especially noticeable in the low-pressure, low-temperature region. This region is further from the LVS and its behavior is less directly affected by pre-spinodal effects, but the inclusion of an explicit LVS in the model is necessary in order to fit it. This is probably because previous attempts to model the behavior at higher temperatures and very low pressures ignored the LVS and relied on polynomial ``background" terms, which led to over-fitting in that region and poor predictions elsewhere. A more theoretically grounded approach to the higher-temperature region, incorporating a LVS, solves this problem. In particular, Ref.~\onlinecite{Singh_2016} implemented the expression for the Gibbs energy of pure state A as a sixth-order polynomial in $\Delta \hat{T}$ and $\Delta \hat{P}$.  Here the terms used for $\{c_{mn}\}$ go only to fourth order, \textit{i. e.} the background expressions for the response functions are quadratic rather than quartic. A comparison between simulation data, TSEOS, and background terms, along selected isobars and isotherms, is provided in Supplementary Figure S2. It shows in particular the effect of the LVS which is included in the background.

\begin{figure}[tttt]
\includegraphics[width = 0.99\columnwidth]{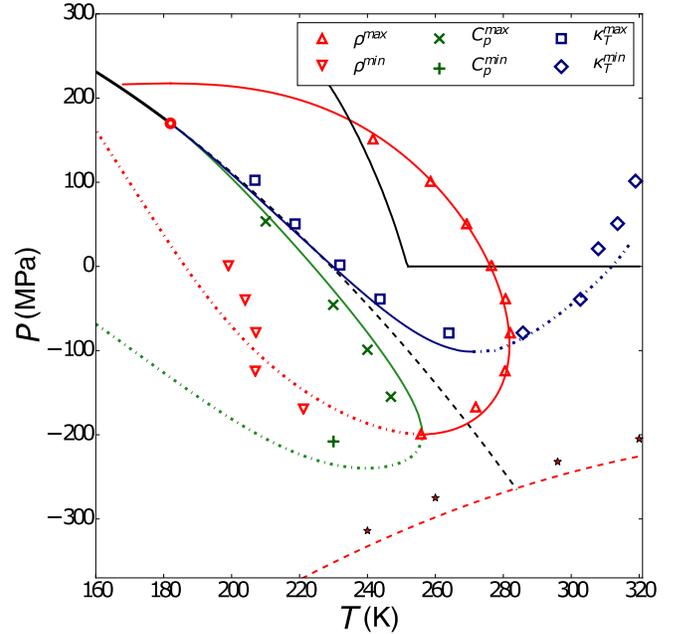}
\caption[Lines of thermodynamic extrema in TIP4P/2005 compared to the TSEOS]{Symbols represent simulation data on the extrema of various thermodynamic properties as shown in the legend, simulated for Ref.~\onlinecite{Gonzalez_2016}. Solid and dot/dashed lines of the same color represent the predictions of the TSEOS for maxima and minima, respectively. The solid black line, open red circle, and dashed black line are the LLPT, LLCP, and Widom line, respectively. The red dashed line is the LVS, and the red stars show the points at which cavitation was observed in simulations. The saturated vapor pressure line and melting line are also represented by solid black lines. Thermodynamic consistency requires that the point at which the locus of density maxima joins the locus of density minima also be an extremum of $C_P$ measured along the isotherm,~\cite{Poole_2005} and that maximum-temperature point on the LDM also be an extremum of $\kappa_T$ measured along the isobar,~\cite{Sastry_1996} as is in fact observed. \label{TDlines}}
\end{figure}
Figures \ref{KtIsobars} to \ref{CpIsobars} show a series of other thermodynamic quantities. The effects of the two key features that this work aims to capture--the LVS and the LLCP--can both be seen clearly in the isothermal compressibility $\kappa_T$ data along isotherms: the compressibility goes through a maximum in the vicinity of the Widom line, decreases and then begins to increase once again as the LVS is approached. This effect is captured by our present extension of the TSEOS, as shown in Fig.~\ref{KtIsotherms}. The isobaric heat capacity $C_P$ is also strongly affected by the presence of the spinodal at low pressures, and is matched well by the TSEOS, with the exception of a few data points at very low $T$ and $P$, as shown in Figs.~\ref{CpIsotherms} and \ref{CpIsobars}. This discrepancy is surprising, because the TSEOS is thermodynamically consistent; therefore, as it reproduces well the temperature/pressure/density relation and the compressibility data, it should reproduce the heat capacity data in the same range as well. For a few of these points, we have carried out longer runs (lasting $500\,\mathrm{ns}$), and these results for the heat capacity are shown in Figs.~\ref{CpIsotherms} and \ref{CpIsobars}. At $-80\,\mathrm{MPa}$ and two intermediate temperatures (260 and $280\,\mathrm{K}$) (Fig.~\ref{CpIsobars}), the difference with shorter runs is barely noticeable and the fit with the TSEOS is good. However, at extreme low temperatures and pressures, the longer runs yield a significantly different value, in better agreement with the TSEOS (Figs.~\ref{CpIsotherms} and \ref{CpIsobars}). This shows that accurate computation of heat capacity at these extreme conditions requires expensive simulations, maybe even longer than what is possible at the present time. To be conservative, we exclude this extreme region from the region of validity of the model (Fig.~\ref{RhoIsochores}).  Figs.~\ref{RhoIsotherms}, \ref{KtIsotherms}, and \ref{CpIsotherms} also include extrapolations of the TSEOS presented in Ref.~\onlinecite{Singh_2016}, showing the improvement of this work over previous formulations at negative pressures.

The most concise presentation of the anomalous thermodynamic behavior of TIP4P/2005 water is shown in Fig.~\ref{TDlines}, which summarizes the lines of minima and maxima in the model's thermodynamic properties. The simulation data can be summarized as follows: the LDM bends to lower temperatures at very low pressures, eventually ending where it meets a locus of minimum density, and loci of finite maxima in $\kappa_T$ and $C_P$ become arbitrarily close to each other upon pressurization and cooling. Qualitatively, our equation of state accounts for this picture in terms of two-structure thermodynamics, with the behavior influenced by the liquid-vapor spinodal. The loci of maxima in the response functions approach each other, together with the critical isochore, at a critical point, where the response functions diverge. At first, the TMD line has a negative slope for the same reason that the LLPT does: lower pressure favors the low-density phase. At very low pressures, however, the effects of the LVS become more significant, pushing the LDM to lower temperatures and eventually forcing it to reach a minimum pressure where it merges with a line of density minima along isobars. Between this minimum pressure and the spinodal pressure, density is monotonic along isobars. Both qualitatively and quantitatively, the match between simulation and theory is remarkable.

The lines of density and response-function extrema and the liquid-vapor spinodal in TIP4P/2005 water demonstrate a pattern which is strikingly similar to that observed in another simulated water-like model, ST2.~\cite{Poole_2005} Using a classical Stillinger-Weber (SW) potential, the cases of water, silicon, and germanium have been investigated:~\cite{Lu_2016,Dhabal_2016} as in TIP4P/2005 water, the TMD line does not touch the LVS, and the latter does not show a retracing behavior. In addition, for SW silicon, a line of compressibility maxima has also been reported.~\cite{Vasisht_2011} However, the classical SW potential may not reliably represent the thermodynamic properties of real silicon. Most recently, Zhao et al.~\cite{Zhao_2016b} revisited the phase behavior of doubly metastable silicon by performing \textit{ab initio} MD simulations. Their results show that the LLPT line in this silicon model goes to deeply negative pressure, until it is terminated by the LLCP at the intersection with the LVS. The LVS, which above the critical temperature was the HDL-vapor spinodal, now becomes the LDL-vapor spinodal that continues to more negative pressures without retracing. There are two other limits of stability within the liquid state, namely the liquid-liquid spinodals: the absolute stability limit of HDL with respect to LDL, and vice-versa. The behavior of this silicon model is equivalent to a “critical-point-free” scenario discussed by several authors.~\cite{Poole_1994,Stokely_2010,Angell_2016,Gallo_2016} In this scenario, there is no line of compressibility maxima, but instead the isothermal compressibility diverges everywhere along the limits of stability. Which scenario is more adequate for real water is still an open question. 

\section{Conclusion}

We present an equation of state that accurately describes the simulation data of TIP4P/2005 over a very wide range of temperatures and pressures including doubly metastable states. The parameters of the TSEOS were obtained by fitting a subset of simulation data, but the resulting equation of state accurately represents the totality of our extensive simulations. This TSEOS might therefore be used as a benchmark to check future simulations with the TIP4P/2005 potential.

Our equation of state accounts for two crucial features: the competition between two interconvertible structures on the one hand, and the liquid-vapor spinodal on the other. Contrary to a theory that attributes the anomalies of supercooled water to a “retracing spinodal”, we find that the observed anomalies in both the density and the response functions in TIP4P/2005 arise as a result of the competition between the two structures, while the liquid-vapor spinodal influences the loci of extrema as shown in Fig.~\ref{TDlines}. Several other models of water exhibit a similar pattern for these characteristic lines. Since the situation in real water remains unresolved, further experimental studies of water in the doubly metastable region are highly desirable.

\section*{Supplementary Material}

See supplementary material for figures showing the fraction of molecules participating in structure B, and simulation data, TSEOS predictions and background terms, along selected isobars and isotherms.

\section*{Acknowledgments}
We acknowledge Vincent Holten, Eduardo Sanz and Carlos Vega for helpful discussions. This work has been funded by grants FIS2013-43209-P, FIS2016-78117-P and FIS2016-78847-P of the Mineco Spain. Chantal Valeriani acknowledges financial support from a Ramon y Cajal tenure track. This work has been possible thanks to a CPU time allocation of the RES (QCM-2015-1-0029 and QCM-2016-1-0036). Mikhail Anisimov acknowledges the Institute of Multiscale Science and Technology (Labex iMUST) supported by the French ANR during part of his sabbatical leave from the University of Maryland, and John Biddle acknowledges a Chateaubriand Fellowship from the Embassy of France in the United States.

\bibliography{PhysRefsv5}

\end{document}